\documentclass[aps,prb,preprint,groupedaddress,showpacs,amsmath,amssymb]{revtex4}

\usepackage{graphicx}
\usepackage{dcolumn}
\usepackage{bm}
\usepackage{hyperref}

\begin{document}

\title{Comment on “Unconventional enhancement of ferromagnetic interactions in Cd-doped GdFe$_2$Zn$_{20}$ single crystals studied by ESR and $^{57}$Fe M\"ossbauer spectroscopies”}

\author{ Paul C. Canfield$^{1,2}$, Sergey L. Bud'ko$^{1,2}$, Andriy Palasyuk$^{1}$, and Tyler J. Slade$^{1,2}$}

\affiliation{$^{1}$Ames Laboratory, US DOE, Iowa State University, Ames, Iowa 50011, USA}
\affiliation{$^{2}$Department of Physics and Astronomy, Iowa State University, Ames, Iowa 50011, USA}

\date{\today}

\begin{abstract}
In the recent publication, Phys. Rev. B {\bf 102}, 144420 (2020), Cabrera-Baez et al. present a study of the effects of Cd-substitution for Zn in the ferromagnetic compound GdFe$_2$Zn$_{20}$.  As part of this paper they claim that for GdFe$_2$Zn$_{18.6}$Cd$_{1.4}$ the effective moment of Gd is reduced by 25\% and the saturated moment of Gd is reduced by over 40\%.  We regrew representative members of the GdFe$_2$Zn$_{20-x}$Cd$_x$ series and did not find any such reductions.  In addition, we measured several crystals from the growth batch that was used by Cabrera-Baez et al. and did not see such reductions.  Although there is a modest increase in $T_C$ with Cd substitution, there is no significant change in the Gd effective moment or the saturated moment associated with the low temperature ferromagnetic state.

\end{abstract}



\maketitle

The {\it RT}$_2$Zn$_{20}$ ({\it R} = rare earth, {\it T} = transition metal) family of materials \cite{nas97a} is remarkable for its versatility and diversity of possible electronic and magnetic ground states.  For {\it R} = Yb, a half dozen, isostructural, heavy fermions were discovered \cite{tor07a,mun12a,can08a};  for {\it R} = Gd – Tm diverse, well defined, local moment behavior has been studied \cite{jia08a,jia09a,jia07a,tia10a,nin11a}; for {\it T} = Fe and {\it R} = Y, Lu, Gd-Tm, greatly enhanced ferromagnetic (or nearly ferromagnetic) behavior has been found \cite{can08a,jia08a,jia09a,jia07a,nin11a,jia07b}.  YFe$_2$Zn$_{20}$ and LuFe$_2$Zn$_{20}$ are nearly ferromagnetic, being closer to the Stoner limit than elemental Pd \cite{can08a,jia07b}.  As such, the {\it R}Fe$_2$Zn$_{20}$ ({\it R} = Gd – Tm) series manifests anomalously large Curie temperatures, $T_C$, relative to the much lower N\'eel temperatures determined for the analogous {\it RT}$_2$Zn$_{20}$ ({\it T} = Co, Rh, Ir) compounds \cite{jia08a,jia09a,tia10a}.  The Stoner enhancement seen for YFe$_2$Zn$_{20}$, LuFe$_2$Zn$_{20}$ and GdFe$_2$Zn$_{20}$ can be systematically and comprehensively tuned by Co-substitution on the Fe site \cite{can08a,jia07b} as well as by Al-substitution on the Zn site  \cite{nin11a}.  For all of these studies, the size of the Gd effective and saturated moment values has remained close to the expected Hund’s rules values: $7.9~\mu_B$ and $7~\mu_B$ respectively.  This is not unusual given that Gd is often the “go-to” rare earth when studying rare earth intermetallic series (e.g. {\it R}BiPt, {\it R}Ni$_2$Ge$_2$, {\it R}Fe$_2$Ge$_2$, {\it R}Co$_2$Ge$_2$, {\it R}AgSb$_2$) \cite{can91a,bud99a,avi04a,kon14a,mye99a} to gain a basic understanding of the magnetic interactions without the potential complications of crystal electric field (CEF) splitting (given that for Gd, $L = 0$) or hybridization (such as can occur for Ce, Yb, or sometimes Sm).  For these reasons, Gd-based compounds are well known for reliably manifesting well defined, isotropic (in the paramagnetic state), local moment behavior with no hybridization or CEF effects. 

\begin{figure}
\begin{center}
\includegraphics[angle=0,width=140mm]{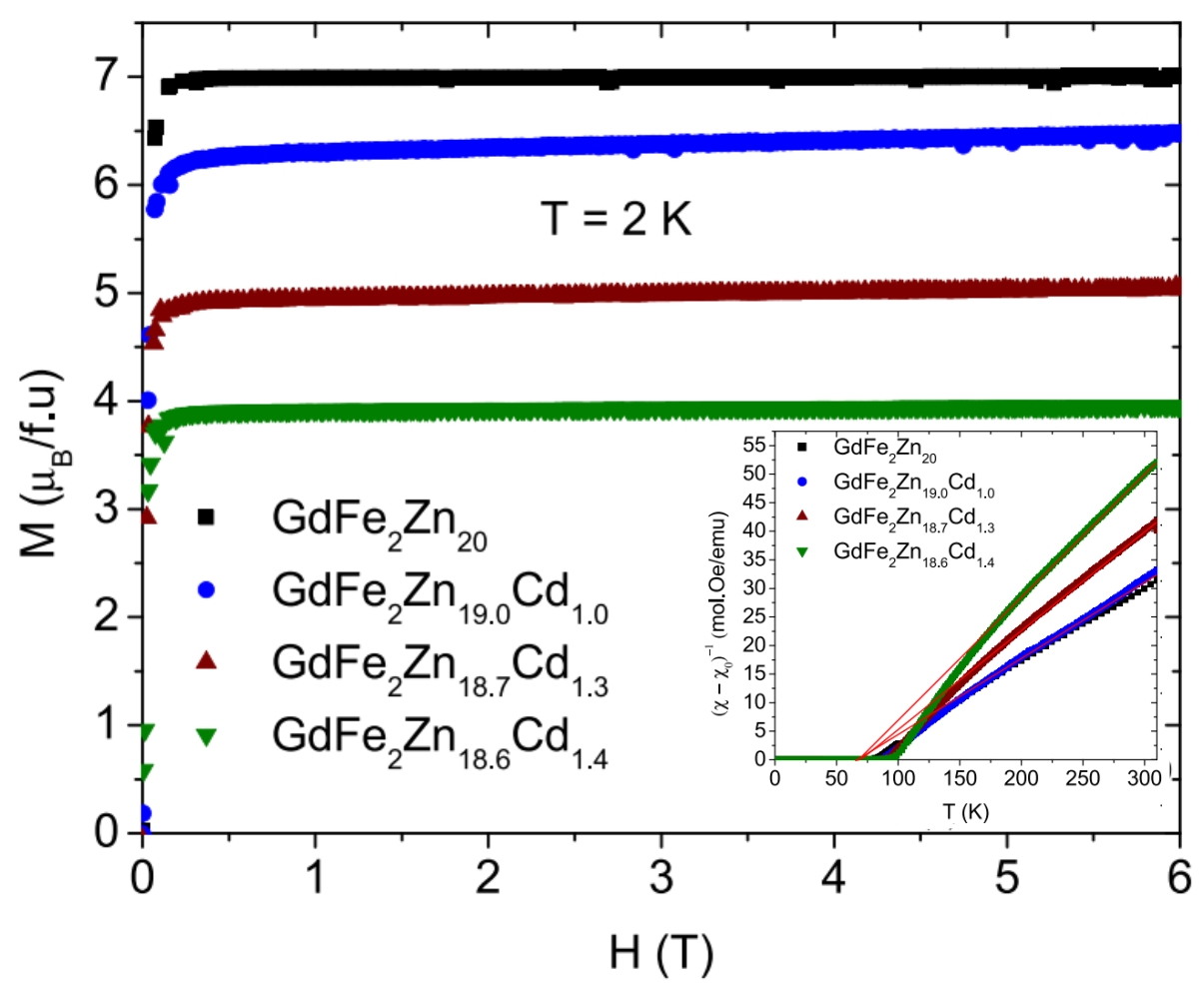}
\end{center}
\caption{(Color online) Field-dependent magnetization $M(H)$ at 2 K for GdFe$_2$Zn$_{20-x}$Cd$_x$ for $0.0  \leq x \leq  1.4$. Inset: inverse magnetic susceptibility, $(\chi - \chi_0)^{-1}$, as a function of temperature for the same samples. $\chi_0$ is the temperature-independent susceptibility component. The figure is based on the figures 3 and 4 from Ref. \onlinecite{cab20a}.  } \label{F1}
\end{figure}

The recent Physical Reviev B publication, “Unconventional enhancement of ferromagnetic interactions in Cd-doped GdFe$_2$Zn$_{20}$ single crystals studied by ESR and $^{57}$Fe M\"ossbauer spectroscopies”  by M. Cabrera-Baez, et al.\cite{cab20a}, caught our attention because, based on the authors’ study of the GdFe$_2$Zn$_{20-x}$Cd$_x$ system, there is a clear claim that there was, “an unexpected increase in $T_C$ for 86 to 96 K, together with a reduction of the magnetic effective moment and saturated magnetic moment”.  Whereas the modest change in $T_C$ is not unexpected, the dramatic reductions of the magnetic effective moment by 25\% and saturated magnetic moment by over 40\% were indeed incredible.  These reductions were inferred, primarily, from temperature and field dependent magnetization data, $M(T)$ and $M(H)$ (see Fig. \ref{F1} which is based on Figs. 3 and 4 from Ref. \onlinecite{cab20a}).  Given our long standing interest in the {\it RT}$_2$Zn$_{20}$ system \cite{tor07a,mun12a,can08a,jia08a,jia09a,jia07a,tia10a,nin11a,jia07b,bud15a} as well as our even longer standing interest in the magnetism of rare earth bearing intermetallic systems \cite{can91a,bud99a,avi04a,kon14a,mye99a} we wanted to understand this potentially unique example of a Gd-based compound having such tunable effective and saturated moments.

We grew single crystals of GdFe$_2$(Zn$_{1-x}$Cd$_x$)$_{20}$ using the growth protocols we developed for the {\it RT}$_2$Zn$_{20}$ system described in Refs. \onlinecite{tor07a,jia07a,jia07b}.  So as to best reproduce the samples described in Ref. \onlinecite{cab20a} we used similar stoichiometries of initial melts.  Table \ref{T1}  presents the elemental analysis from energy dispersive spectroscopy (EDS) for our four growth as well as the lattice parameter, $a$, determined from powder x-ray diffraction.  As can be seen, the Cd substitutes for the Zn in a monotonic manner with a systematic, slight, increase of the unit cell dimension.  So as to allow for more direct comparison of results, we will use the GdFe$_2$Zn$_{20-x}$Cd$_x$ notation adopted in Ref. \onlinecite{cab20a}.  The EDS analysis of our samples showed that we grew $x = 0, 0.71, 1.06$ and 1.52.  The data presented in Table \ref{T1} are in basic agreement with data presented in  Ref. \onlinecite{cab20a}, strongly suggesting that the samples are similar.  Although evaluation of Curie temperatures was not the focus of our study, we were able to infer an $\sim 7$ K increase in $T_C$ as Cd substitution was increased, again consistent with  Ref. \onlinecite{cab20a}.

\begin{figure}
\begin{center}
\includegraphics[angle=0,width=140mm]{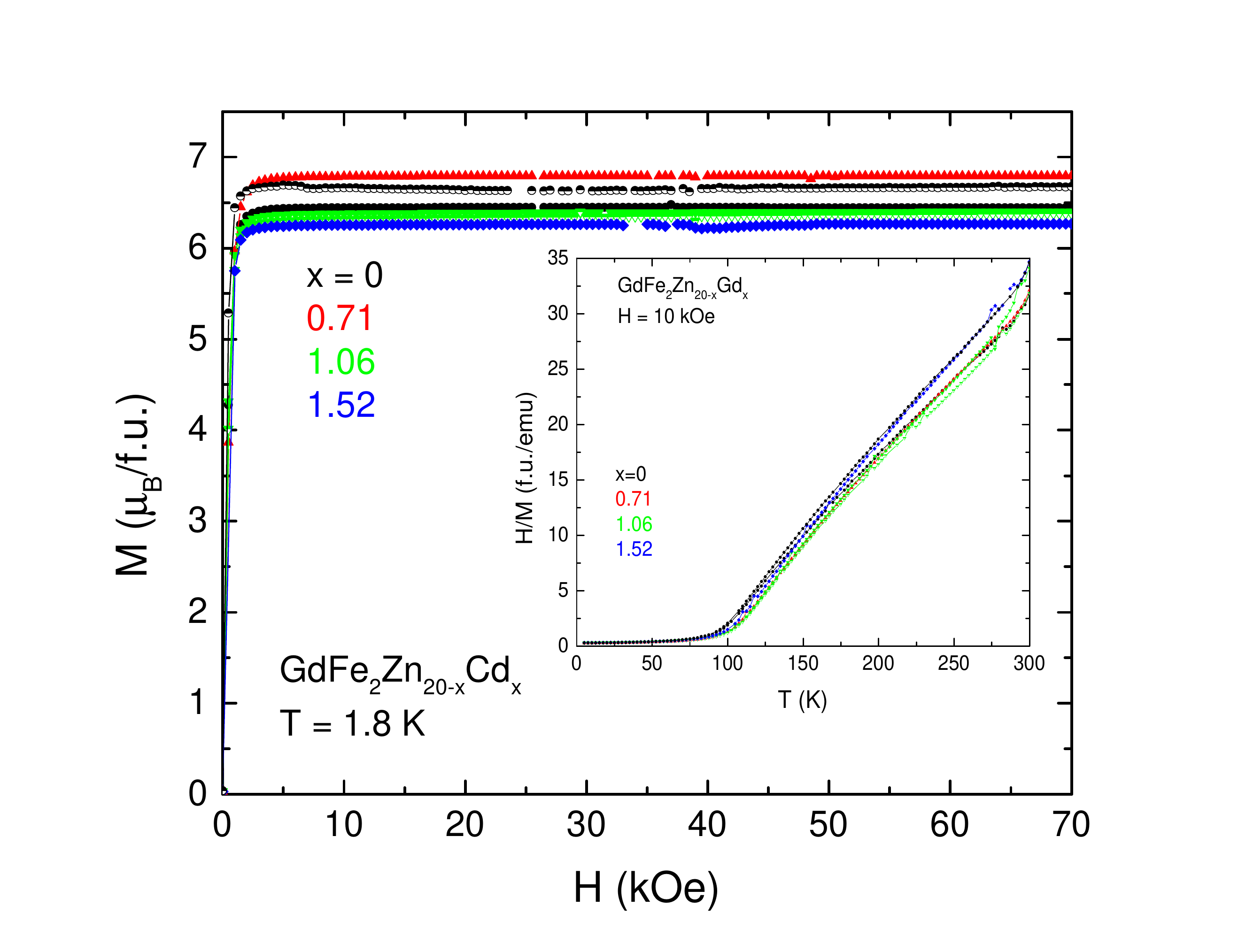}
\end{center}
\caption{(Color online) Field-dependent magnetization $M(H)$ at 1.8 K for GdFe$_2$Zn$_{20-x}$Cd$_x$ for Ames samples. Inset: inverse magnetic susceptibility, $H/M$, measured in $H = 10$~kOe as a function of temperature for the same samples.} \label{F2}
\end{figure}

Figure \ref{F2} presents the $M(H)$ data for $T = 1.8$~ K and the $H/M(T)$ data for $H = 10$~ kOe for $x = 0, 0.72, 1.06$ and 1.52 Cd substitution.  These two data sets are conspicuously different from the data presented in Ref. \onlinecite{cab20a} and reproduced in Fig. \ref{F1}.  These data show that there is little or no systematic change in either the effective or saturated moment of the GdFe$_2$Zn$_{20-x}$Cd$_x$ compounds as Cd in increased from $x = 0$ to 1.5.  This result is consistent with the notion that Gd has a robust effective and saturated moment size and is not susceptible to CEF splitting or hybridization.  This result is also qualitatively consistent with the specific heat data presented in Fig. 2a of Ref. \onlinecite{cab20a} which show no qualitative change in the specific heat anomaly (or associated entropy change) as $x$-increases.  At a more quantitative level, this result is also consistent with the magnetic hyperfine field data measured with $^{57}$Fe Mossbauer spectroscopy (Fig. 6 of Ref. \onlinecite{cab20a}) where the $T = 3$~ K hyperfine field is identical for $x = 0$ and $x = 1.4$ samples.  Whereas this is very unlikely to happen if the $x = 1.4$ compound’s saturated moment is reduced by over 40\%, it is expected if the $x = 0$ and $x = 1.4$ compounds have saturated moments that are virtually identical.

\begin{figure}
\begin{center}
\includegraphics[angle=0,width=140mm]{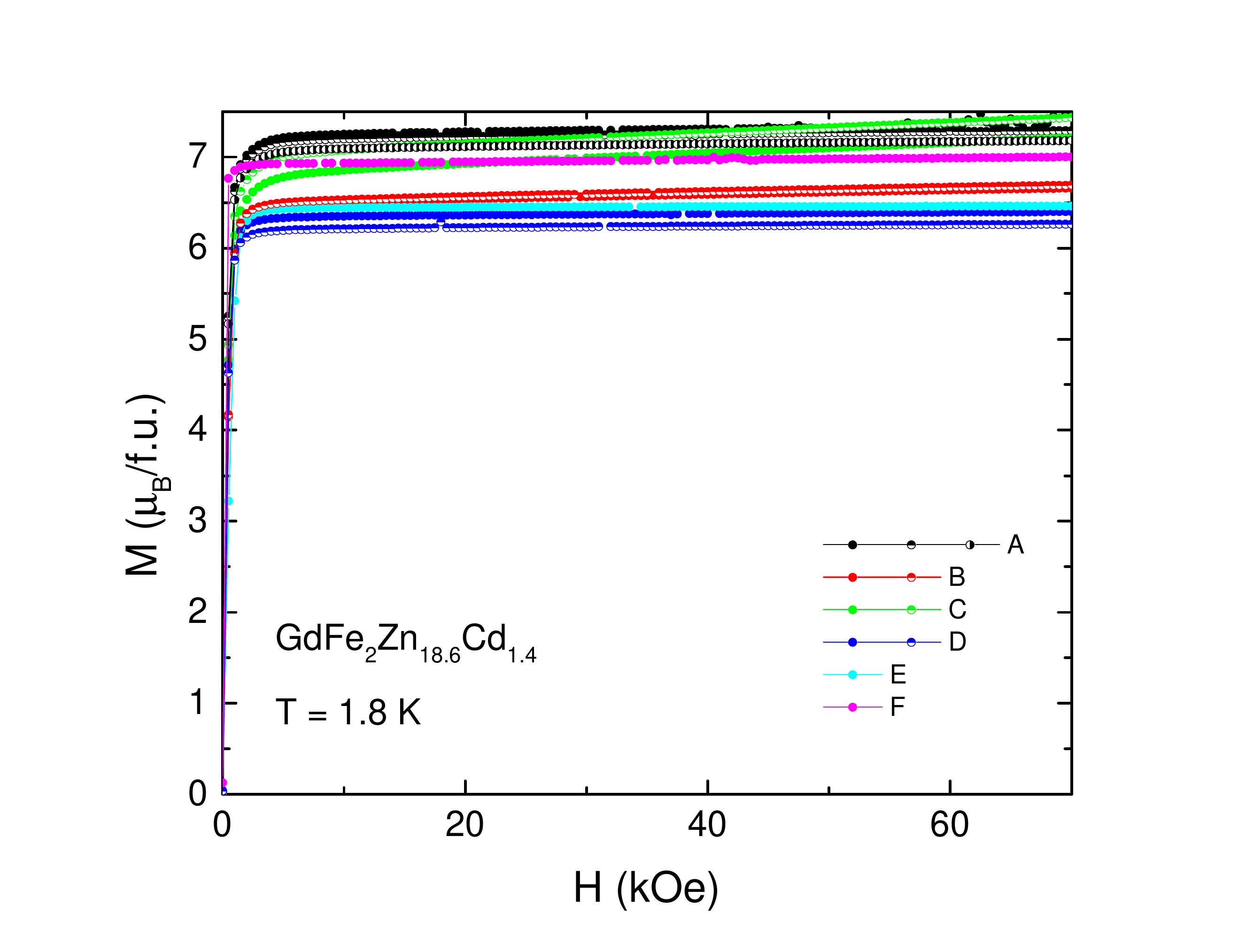}
\end{center}
\caption{(Color online) Field-dependent magnetization $M(H)$ at 1.8 K for GdFe$_2$Zn$_{18.6}$Cd$_{1.4}$ samples from the same batch \cite{rema} as those used in Ref. \onlinecite{cab20a}. Note, multiple samples were measured more than once as marked by different symbols in the legend.} \label{F3}
\end{figure}

So as to further resolve the clear discrepancy between our data (Fig. \ref{F2}) and the data presented in Ref. \onlinecite{cab20a}, we were able to measures samples from the same batch of crystals that the GdFe$_2$Zn$_{18.6}$Cd$_{1.4}$ samples used in  Ref. \onlinecite{cab20a} came from \cite{rema}.  We  measured $M(H)$ at 1.8~K on six different crystals (Fig. \ref{F3}).  As can be seen the $M(H)$ data have $\mu_{sat}$ values between 6 and 7 $\mu_B$ /f.u., consistent with the $\mu_{sat}$ values we found for our GdFe$_2$Zn$_{20-x}$Cd$_x$ series (Fig. \ref{F2}) and inconsistent with the data presented in  Ref. \onlinecite{cab20a} (shown in Fig. \ref{F1} above).  For completeness, we used EDS to determine the compositions of the two crystals with extreme values of $\mu_{sat}$ (labeled A and D) and include these data at the bottom of Table \ref{T1}.  As can be seen the Cd substitution level of these samples is comparable to (albeit slightly larger than) our $x = 1.52$ sample.

Given that these samples are grown out of excess Zn, it is possible that the specific samples used for $M(T)$ and $M(H)$ measurements in Ref. \onlinecite{cab20a} may have had some excess Zn on/in them.  Since Zn is non-magnetic, this would lead to an apparent decrease in the effective and saturated moments inferred for the samples.  This would also be consistent with the need to subtract off non-trivial $\chi_0$ terms when analyzing the effective moments (as was done in Fig. 3 of Ref. \onlinecite{cab20a}).  It should also be noted that, based on the analysis of the data above, it seem likely that a similar error was made for the $x = 1.3$ data presented in Ref. \onlinecite{cab20a} since the $\mu_{sat}$ value for this sample is also well below the values we found for our doping series (Fig. \ref{F2}) or the measurements we made on the crystals associated with ref. [16] (Fig. 3).

All in all we have to conclude that the data shown in Figs. 3 and 4 of Ref. \onlinecite{cab20a} (reproduced in Fig. \ref{F1} above) are incorrect and that the conclusions drawn from these data are also incorrect.  There is no dramatic reduction of the Gd effective or saturated moment as Cd is substituted for Zn in the GdFe$_2$Zn$_{20-x}$Cd$_x$ series.  Given that the Gd moment is not changing in any significant manner with Cd substitution for Zn, the modest change in $T_C$ is not unexpected either.  For the GdFe$_2$(Zn$_{1-x}$Al$_x$)$_{20}$ series $T_C$ changes from 86 K (for $x = 0$) to 10 K (for $x = 0.122$); for the Gd(Fe$_{1-x}$Co$_x$)$_2$Zn$_{20}$  a $T_C$ of 86~ K (for $x = 0$) changes to a $T_N$ of 5.7~ K (for $x = 1.0$).  These changes have been very clearly correlated with changes in how close the electronic system is to the Stoner limit, with the addition of extra electrons to the conduction band (via Co- or Al-substitutions) moving the system away from this limit.  Cd, on the other hand, is iso-electronic, so such dramatic changes in $T_C$ are not anticipated.  An increase in $T_C$ with increasing lattice parameter is anticipated, though, based on hydrostatic pressure studies of pure GdFe$_2$Zn$_{20}$ which showed that $T_C$ decreased by $\sim 1.2$~ K under 7~ kbar of pressure \cite{jia08a}.  This result implies that negative chemical pressure (i.e. increases in the lattice parameter such as are seen with Cd substitution) would be expected to lead to increases in $T_C$. Using calculated bulk modulus for YFe$_2$Zn$_{20}$, $K = 1.48$~ Mbar \cite{che11a}, relative volume change $\Delta V/V_0 \approx 0.016$ between $x = 0$ and $x = 1.52$ from the lattice parameters in Table \ref{T1},  measured \cite{jia08a} pressure derivative $dT_C/dP = - 0.17$~ K/kbar, and assuming equivalence of physical and chemical pressure, we evaluate expected increase in $T_C$ of $\sim 4$~ K. Despite the oversimplified assumptions, this estimate is within the factor of 2 of the observed $\Delta T_C \sim 7$~ K.

In summary, whereas Cd substitution for Zn in GdFe$_2$Zn$_{20-x}$Cd$_x$ does lead to a modest increase in the ferromagnetic Curie temperature, $T_C$, the size of the Gd effective and saturated moments is unaffected as $x$ varies from 0 to 1.5.

\begin{acknowledgments}

We thank Michael Cabrera-Baez who kindly agreed to supply us with the  remains of the growth that was identified as GdFe$_2$Zn$_{18.6}$Cd$_{1.4}$ in Ref. \onlinecite{cab20a} for the cross-check measurements. Useful discussions with Dominic H. Ryan are appreciated. Work at the Ames Laboratory was supported by the U.S. Department of Energy, Office of Science, Basic Energy Sciences, Materials Sciences and Engineering Division. The Ames Laboratory is operated for the U.S. Department of Energy by Iowa State University under contract No. DE-AC02-07CH11358. A. P.  was supported by the Critical Materials Institute, an Energy Innovation Hub funded by the U.S. Department of Energy, Office  of  Energy  Efficiency  and  Renewable  Energy,  Advanced Manufacturing Office. T.J.S. was supported by by the Gordon and Betty Moore Foundation’s EPiQS Initiative through Grant No. GBMF4411.
\end{acknowledgments}

\clearpage

\begin{table}

\caption{Results of EDS analysis of four Ames samples and two samples identified as  GdFe$_2$Zn$_{18.6}$Cd$_{1.4}$ in Ref. \onlinecite{cab20a}. The columns are: atomic percents of Gd, Fe, Zn, and Cd from EDS analysis, the value of $x$ in the GdFe$_2$Zn$_{20-x}$Cd$_x$ notation, and the lattice parameter $a$ for Ames samples. } \label{T1}

\begin{tabular}{ c | c c c c c c  }

  Sample~ &~ Gd &~ Fe &~ Zn &~ Cd &~ $x$ &~ $a$ (\AA)\\
\hline
 Ames 1~ &~ 4.49 &~ 8.71 &~ 86.7 &~ 0 &~ 0 &~ 14.125 \\
 Ames 2~ &~ 4.42 &~ 8.83 &~ 83.66 &~ 3.08 &~ 0.71 &~ 14.159 \\
 Ames 3~ &~ 4.32 &~ 8.79 &~ 81.88 &~ 4.58 &~ 1.06 &~ 14.179 \\
 Ames 4~ &~ 4.43 &~ 8.78 &~ 79.90 &~ 6.55 &~ 1.52 &~ 14.200 \\
\hline
Ref. \onlinecite{cab20a} A~ &~ 4.46 &~ 8.87 &~ 79.09 &~ 7.58 &~ 1.74 \\
Ref. \onlinecite{cab20a} D~ &~ 4.52 &~ 8.92 &~ 78.79 &~ 7.77 &~ 1.80 \\

\end{tabular}

\end{table}

\clearpage

\clearpage

\clearpage

\begin{thebibliography}{99}
\bibitem{rema}
Given our results, we contacted the authors of Ref. \onlinecite{cab20a} and M. Cabrera-Baez sent us the remains of the growth that was identified as GdFe$_2$Zn$_{18.6}$Cd$_{1.4}$.  Unfortunately the exact sample that was measured for Ref. \onlinecite{cab20a} was no longer available; we were able to remove and measure six different samples from this batch.
\end{thebibliography}
\end{document}